\documentstyle[ppmo209,twoside,epsf,fancyhdr]{article}
\input{psfig.sty}                               
\def\apjs{{ApJS}}

\def\hst{{\it HST}}

\def\lax{{$\mathrel{\hbox{\rlap{\hbox{\lower4pt\hbox{$\sim$}}}\hbox{$<$}}}$}}
\def\gax{{$\mathrel{\hbox{\rlap{\hbox{\lower4pt\hbox{$\sim$}}}\hbox{$>$}}}$}}
\def\simlt{\lower.5ex\hbox{$\; \buildrel < \over \sim \;$}}
\def\simgt{\lower.5ex\hbox{$\; \buildrel > \over \sim \;$}}

\def\percm2{cm$^{-2}$}

\begin{document}

\markboth {Sylvain Veilleux} {Local ULIRGs and Quasars}

\title{Local Ultraluminous Infrared Galaxies and Quasars}

\author{Sylvain Veilleux$^{1}$}

\inst{$^1$ Department of Astronomy, University of Maryland, College
  Park, MD 20742, USA}
\email{veilleux@astro.umd.edu}

\date{Received~2006 February 5; accepted~2006 February 5}

\begin{abstract}
  \ This paper reviews the recent results from a comprehensive
  investigation of the most luminous mergers in the local universe,
  the ultraluminous infrared galaxies (ULIRGs) and the quasars.
  First, the frequency of occurrence and importance of black hole
  driven nuclear activity in ULIRGs are discussed using the latest
  sets of optical, near-infrared, mid-infrared, and X-ray spectra on
  these objects. Obvious trends with luminosity, infrared color, and
  morphology are pointed out.  Next, the host galaxy properties of
  ULIRGs are described in detail and then compared with local quasar
  hosts and inactive spheroids. By and large, these data are
  consistent with the scenario where ULIRGs are intermediate-mass
  elliptical galaxies in formation and in the process of becoming
  moderate-luminosity optical quasars. The powerful galactic winds
  detected in many ULIRGs may help shed any excess gas during this
  transformation. However, this evolutionary scenario does not seem to
  apply to all ULIRGs and quasars: Ultraluminous infrared mergers do
  not always result in a quasar, and low-luminosity quasars near the
  boundary with Seyferts do not all show signs of a recent major
  merger.

\keywords{galaxies: active --- galaxies: interactions --- galaxies:
  Seyfert --- galaxies: starburst --- infrared: galaxies}

\end{abstract}

\section{Introduction}           

Galaxy merging is a key driving force of galaxy evolution. In
hierarchical cold dark matter models of galaxy formation and
evolution, merging leads to the formation of elliptical galaxies,
triggers major starbursts, and may account for the growth of
supermassive black holes and the formation of quasars (e.g., Kauffmann
\& Haehnelt 2000). The importance of mergers increases with redshift
(e.g., Zepf \& Koo 1989; Carlberg, Pritchet, \& Infante 1994;
Neuschaefer et al. 1997; Khochfar \& Burkert 2001). It is clear that
dust-enshrouded starbursts and active galactic nuclei (AGN) play an
extremely important role in the high-redshift Universe and are
probably the dominant contributors to the far-infrared/submm and X-ray
backgrounds, respectively (e.g., Pei, Fall, \& Hauser 1999; Miyaji,
Hasinger, \& Schmidt 2000).  These luminous, merger-induced starbursts
and AGN at high redshift thus provide readily observable signposts for
tracing out the main epoch of elliptical galaxy and quasar formation
if the above scenario is correct.

In order to assess quantitatively the physics of the merger process
and its link to the epoch of elliptical and QSO formation at high
redshift we must first understand the details of galaxy merging and
its relationship to starbursts and AGN in the local universe. The most
violent local mergers and the probable analogs to luminous
high-redshift mergers are the ultraluminous infrared galaxies
(ULIRGs). ULIRGs are advanced mergers of gas-rich, disk galaxies
sampling the entire Toomre merger sequence beyond the first
peri-passage (Veilleux, Kim, \& Sanders 2002; hereafter referred as
VKS02 and discussed in more detail in \S 3). ULIRGs are among the most
luminous objects in the local universe, with both their luminosities
($\ge$ 10$^{12}$ L$_\odot$ emerging mainly in the far-IR) and space
densities similar to those of quasars (e.g., Sanders \& Mirabel 1996).
The near-infrared light distributions in many ULIRGs appear to fit a
de Vaucouleurs law (Scoville et al. 2000; VKS02; see \S 3).  ULIRGs
have a large molecular gas concentration in their central kpc regions
(e.g., Downes \& Solomon 1998) with densities comparable to stellar
densities in ellipticals.  These large central gas concentrations (and
stars efficiently forming from them) may be the key ingredient for
overcoming the fundamental phase space density constraints that would
otherwise prevent formation of dense ellipticals from much lower
density disk systems (Gunn 1987; Hernquist, Spergel, \& Heyl 1993).
Kormendy \& Sanders (1992) have proposed that ULIRGs evolve into
ellipticals through merger induced dissipative collapse.  In this
scenario, these mergers first go through a luminous starburst phase,
followed by a dust-enshrouded AGN phase, and finally evolve into
optically bright, `naked' QSOs once they either consume or shed their
shells of gas and dust (Sanders et al. 1988a).

Gradual changes in the far-infrared spectral energy distributions
between `cool' ULIRGs ({\em IRAS} 25-to-60 $\mu$m flux ratio,
$f_{25}/f_{60} < 0.2$), `warm' ULIRGs, and QSOs (Sanders et al. 1988b;
Haas et al. 2003) bring qualitative support to an evolutionary
connection between these various classes of objects, but key elements
remain to be tested. This review attempts to summarize the latest
results from our multiwavelength study of these objects. In \S 2, I
discuss the important issue of nuclear activity in ULIRGs. In \S 3,
the morphological properties of ULIRGs and quasars are compared and
tested against the predictions of the merger-driven evolutionary
scenario. Galactic winds in ULIRGs are discussed in \S 4; I argue
that these winds constitute a key ingredient in this evolutionary
picture. The main conclusions are summarized in \S 5.

\section{Nuclear Activity in ULIRGs}

Considerable effort has been invested in recent years to determine the
frequency of occurrence of AGN in ULIRGs and the importance of nuclear
activity in powering the large infrared luminosities of these objects.
Optical spectroscopy has for many years revealed a trend with infrared
luminosity: only $\sim$ 5\% of all infrared galaxies with log~[$L_{\rm
  IR}$/$L_\odot$] = 10 -- 11 show optical signs of Seyfert activity,
while this fraction reaches $\sim$ 50\% among ULIRGs with log~[$L_{\rm
  IR}$/$L_\odot$] $\ge$ 12.3 (e.g., Veilleux et al. 1995; Kim, Veilleux,
\& Sanders 1998; Veilleux, Kim, \& Sanders 1999; Kewley et al. 2001).

These numbers should be considered lower limits since emission from
circumnuclear starbursts often dilutes the AGN emission-line
signatures in ULIRGs.  The effects of dilution were nicely
demonstrated in a recent \hst/STIS study of four `warm' ULIRGs by
Farrah et al. (2005).  An AGN is detected in the \hst\ data of each of
these objects, while only two objects show signs of an AGN from the
ground.  The STIS spectra of one of these objects (F05189--2524)
present the strongest high ionization lines ever observed in a galaxy.

So far, hard X-ray studies have had only moderate success detecting
buried AGN in ULIRGs (e.g., Ptak et al. 2003; Franceschini et al.
2003). In a recent {\em CXO} study of 14 ULIRGs by our group (Teng et
al. 2005), we found clear trends of increasing $L_{\rm 2 - 10
  keV}/L_{\rm FIR}$ ratio with increasing {\em IRAS} 25-to-60 $\mu$m
flux ratio (Fig. 1). But in most of these objects, our data are not of
sufficiently good quality to distinguish between starburst emission
and emission from a Compton-thick AGN.  Observatories capable of
probing energies above 10 keV, such as {\em Suzaku}, should be able in
the near future to discriminate between these two possibilities.

Extinction due to dust is less important at long wavelengths, so not
surprisingly this is where most of the progress in our understanding
of the ULIRG power source has taken place in recent years.  By and
large, the results from near-infrared and mid-infrared spectroscopic
studies of ULIRGs are consistent with those found at optical
wavelengths. Near-infrared signatures of AGN are found in most
optically-classified Seyfert ULIRGs, but not in LINER or HII ULIRGs
(e.g., Veilleux, Sanders, \& Kim 1997, 1999). Most ULIRGs with an
optical (e.g., H$\beta$, H$\alpha$) or near-infrared (e.g., Pa$\beta$,
Pa$\alpha$, Br$\gamma$) broad-line region are AGN dominated based on
the value of the BLR-to-bolometric luminosity ratio, which is found to
be typical of that of optical quasars.  Similarly, one finds 
excellent agreement between the optical/near-infrared classification
of ULIRGs and the classification based on the equivalent width of the
mid-infrared PAH feature (Genzel et al. 1998; Lutz et al. 1998; Lutz,
Veilleux, \& Genzel 1999; Tran et al. 2001).  This indicates that
strong AGN activity, once triggered, quickly breaks the obscuring
screen at least in certain directions, thus becoming detectable over a
wide wavelength range.

The advent of the {\em Spitzer Space Telescope} ({\em SST}) is
adding considerably to our knowledge on these objects. IRS spectra
have been published for only a few objects so far (e.g., Armus et al.
2004), but the analysis of IRS spectra on several more objects is in
progress. Our group has been allocated 95.3 hours to study 54 local
ULIRGs and QSOs with IRS, but it is too early at this time to discuss
the results from our analysis.

\vskip 0.3cm

\begin{figure}
\hskip +1.2in
\psfig{file=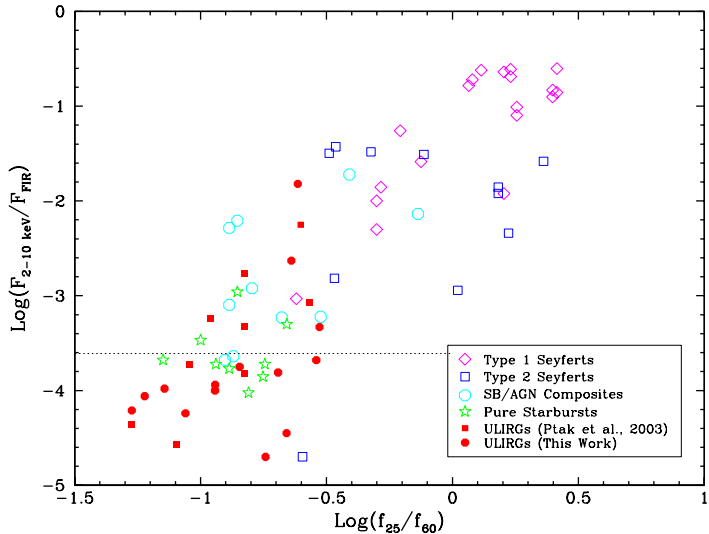,width=4in,angle=-90}
\caption{ Plot of log($f_{\rm 2-10 keV}/f_{\rm FIR}$) vs.  {\em IRAS}
  log($f_{25}/f_{60}$) from Teng et al. (2005). All objects in the
  sample of Teng et al. (2005) are ULIRGs from the 1-Jy sample of Kim
  \& Sanders (1998). The Seyfert 1 ULIRGs are distributed near the
  optical Seyferts, while the other ULIRGs are located among the
  starbursts and composites.  Here we have only included the values
  for the Ptak et al. (2003) ULIRGs derived from their global spectra.
  The dotted line represents the average log($f_{\rm 2-10 keV}/f_{\rm
    FIR}$) values for the pure starbursts.  }
\end{figure}
\vskip 0.6cm

\section{Host Properties}

It is important to compare the results discussed in the previous
section with the host properties of ULIRGs and QSOs. If QSOs are
indeed the end-products of ultraluminous infrared mergers, one should
expect clear trends with morphology and expect the host
luminosities/masses to be similar for both classes of objects.

\subsection{Ground-Based Imaging Study}

In a first attempt to address these questions, we (VKS02)
carried out a systematic R and K$^\prime$ imaging study of the {\em
  IRAS} 1 Jy sample of 118 ULIRGs. This large homogeneous sample
allowed us for the first time to draw statistically meaningful
conclusions; the problems of small sample size and/or inhomogeneous
selection criteria have plagued many studies of luminous infrared
galaxies in the past.

In VKS02, we find that all but one object in the 1-Jy sample show
signs of a strong tidal interaction/merger.  Multiple mergers
involving more than two galaxies are seen in no more than 5 of the 118
($<$ 5\%) systems.  None of the 1-Jy sources is in the first-approach
stage of the interaction, and most (56\%) of them harbor a single
disturbed nucleus and are therefore in the later stages of a merger.
Seyfert galaxies (especially those of type 1), warm ULIRGs
($f_{25}/f_{60} \ge 0.2$) and the more luminous systems ($>$
10$^{12.5}$ $L_\odot$) all show a strong tendency to be advanced
mergers with a single nucleus.

The individual galaxies in the binary systems of the 1-Jy sample show
a broad distribution in host magnitudes (luminosities) with a mean of
--21.02 $\pm$ 0.76 mag. (0.85 $\pm$ $^{0.86}_{0.43}$ $L^\ast$) at $R$
and --23.98 $\pm$ 1.25 mag. (0.90 $\pm$ $^{1.94}_{0.61}$ $L^\ast$) at
$K^\prime$, and a $R$- or $K^\prime$-band luminosity ratio generally
less than $\sim$ 4.  Single-nucleus ULIRGs also show a broad
distribution in host magnitudes (luminosities) with an average of
--21.77 $\pm$ 0.92 mag. (1.69 $\pm$ $^{2.25}_{0.97}$ $L^\ast$) at $R$
and --25.03 $\pm$ 0.94 mag. (2.36 $\pm$ $^{3.24}_{1.38}$) at
$K^\prime$.  These distributions overlap considerably with those of
quasars. 


An analysis of the surface brightness profiles of the host galaxies in
single-nucleus sources reveals that about 73\% of the $R$ and
$K^\prime$ surface brightness profiles are fit adequately by an
elliptical-like de Vaucouleurs law.  These elliptical-like 1-Jy
systems have luminosity and $R$-band axial ratio distributions that
are similar to those of normal (inactive) intermediate-luminosity
ellipticals and follow with some scatter the same $\mu_e - r_e$
relation, giving credence to the idea that some of these objects may
eventually become intermediate-luminosity elliptical galaxies if they
get rid of their excess gas or transform this gas into stars.  These
elliptical-like hosts are most common among merger remnants with
Seyfert 1 nuclei (83\%), Seyfert 2 optical characteristics (69\%) or
mid-infrared ($ISO$) AGN signatures (80\%).  The mean half-light
radius of these ULIRGs is $4.80 \pm 1.37$ kpc at $R$ and 3.48 $\pm$
1.39 kpc at $K^\prime$, typical of intermediate-luminosity
ellipticals. These values are in excellent agreement with the
measurements of McLeod \& McLeod (2001) and Surace, Sanders, \& Evans
(2001) obtained for moderate-luminosity quasars but systematically
lower than the measurements of Dunlop et al.  (2003) obtained for
higher luminosity quasars. I return to this point in \S 3.2.

In general, the results from VKS02 are consistent
with the merger-driven evolutionary sequence ``cool ULIRGs
$\rightarrow$ warm ULIRGs $\rightarrow$ quasars.'' However, many
exceptions appear to exist to this simple picture (e.g., 46\% of the
41 advanced mergers show no obvious signs of Seyfert activity).

\vskip 0.3cm

\begin{figure}
\hskip +2.0in
\psfig{file=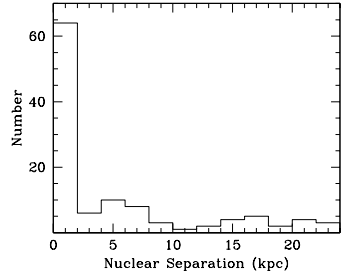,width=2.0in,angle=0}
\caption{ Apparent nuclear separations in the 1-Jy sample of
  ULIRGs. The distribution is highly peaked at small values but also
  presents a significant tail at high values (taken from VKS02). }
\end{figure}
\vskip 0.6cm

\subsection{Recent HST Study}

\vskip 0.3cm

\begin{figure}
\hskip +1.2in
\psfig{file=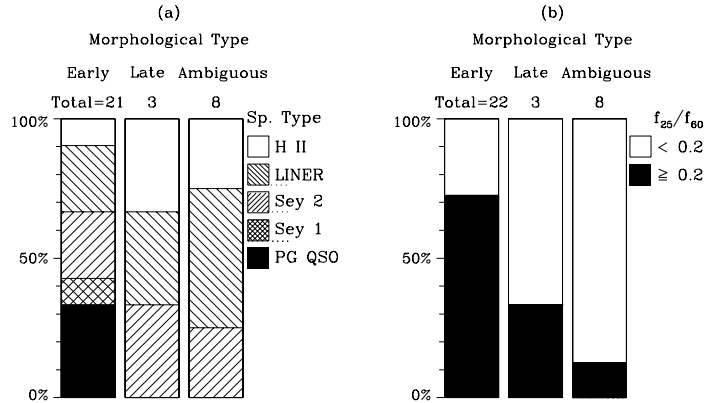,width=4in,angle=0}
\caption{  Trends between the {\em dominant} morphological
  types (based on a decomposition that uses a single S\'ersic
  galaxy component) and ($a$) optical spectral types, ($b$) {\em IRAS}
  25-to-60 $\mu$m colors. There is one fewer object in panel ($a$)
  than in panel ($b$) because the optical spectral type of
  F02021$-$2103 is unknown. The hosts of warm, quasar-like objects all
  have a prominent early-type spheroidal component.  F05189--2524 is
  the only Seyfert 2 ULIRG in the sample with a dominant late-type
  morphology (Veilleux et al. 2006).}
\end{figure}
\vskip 0.6cm

The removal of the central PSF emission associated with the AGN or
nuclear starburst is an important source of errors in the analysis of
the surface brightness profiles in the more nucleated systems of
VSK02. In an effort to verify these results, we (Veilleux et al.
2006) have recently obtained and analyzed deep {\em NICMOS} H-band
images of 26 highly nucleated 1-Jy ULIRGs and 7 IR-excess (L$_{\rm
  IR}$/L$_{\rm BOL}$ $>$ 0.4) PG QSOs.

A detailed two-dimensional analysis of the surface brightness
distributions in these objects confirms that the great majority (81\%)
of the single-nucleus systems show a prominent early-type morphology.
However, low-surface-brightness exponential disks are detected on
large scale in at least 4 of these sources.  The hosts of 'warm'
($f_{25}/f_{60} > 0.2$), AGN-like systems are of early type and have
less pronounced merger-induced morphological anomalies than the hosts
of cool systems with LINER or HII region-like nuclear optical spectral
types (Fig. 3).  The host sizes and luminosities of the 7 PG~QSOs in
our sample are statistically indistinguishable from those of the ULIRG
hosts.  In comparison, highly luminous quasars, such as those studied
by Dunlop et al. (2003), have hosts which are larger and more
luminous.

As shown in Fig. 4, the hosts of ULIRGs and PG~QSOs lie close to the
locations of intermediate-size ($\sim$ 1 -- 2 $L^*$) spheroids in the
photometric projection of the fundamental plane of ellipticals,
although there is a tendency in our sample for the ULIRGs with small
hosts to be brighter than normal spheroids.  Excess emission from a
merger-triggered burst of star formation in the ULIRG/QSO hosts may be
at the origin of this difference. Our results provide support for a
possible merger-driven evolutionary connection between cool ULIRGs,
warm ULIRGs, and PG~QSOs. However, this sequence may break down at low
luminosity since the lowest luminosity PG~QSOs in our sample show
distinct disk components which preclude major (1:1 -- 2:1) mergers.
The black hole masses derived from the galaxy host luminosities imply
sub-Eddington accretion rates for all objects in the sample.

\vskip 0.3cm

\begin{figure}
\hskip +0.6in
\psfig{file=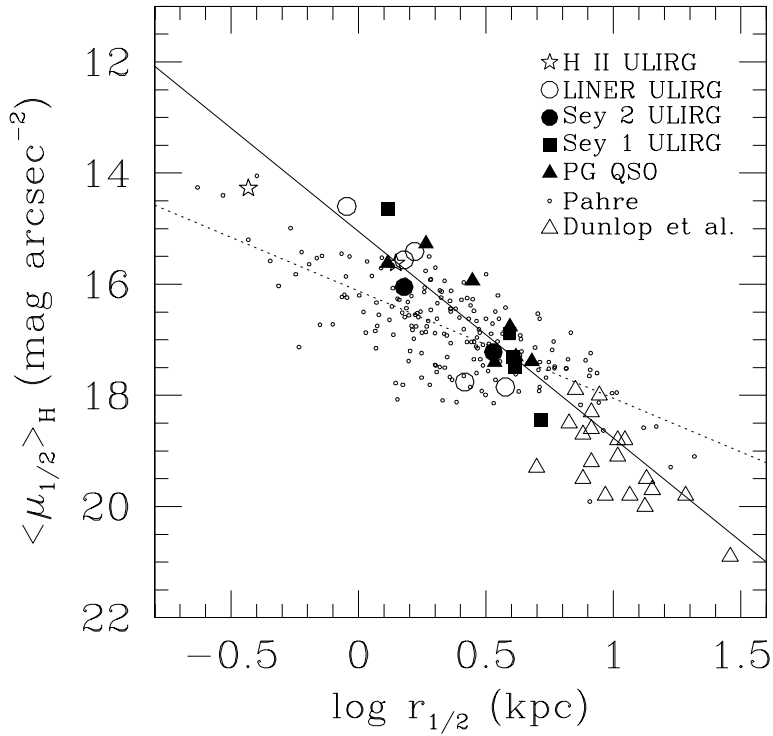,width=5in,angle=0}
\caption{Surface brightness versus half-light radius for the
  early-type host galaxies in the {\em HST} sample of Veilleux et al.
  (2006).  The hosts of the 7 PG~QSOs in our sample are statistically
  indistinguishable from the hosts of the 1-Jy ULIRGs. Both classes of
  objects fall near, but not quite on, the photometric fundamental
  plane relation of ellipticals as traced by the data of Pahre (1999;
  dashed line); the smaller objects in our sample tend to lie above
  this relation (the solid line is a linear fit through our data
  points).  This excess H-band emission may be due to a
  merger-triggered burst of star formation. ULIRGs and PG~QSOs
  populate the region of the photometric fundamental plane of
  intermediate-size ($\sim$ 1 -- 2 $L^*$) elliptical/lenticular
  galaxies.  In contrast, the hosts of the luminous quasars of Dunlop
  et al.  (2003) are massive ellipticals which are significantly
  larger than the hosts of ULIRGs and PG~QSOs.  For this comparison,
  the R-band half-light radii tabulated in Dunlop et al. were taken at
  face value, and the surface brightnesses in that paper were shifted
  assuming R -- H = 2.9, which is typical for early-type systems at
  $z$ $\sim$ 0.2 (see Veilleux et al. 2006 for more detail). }
\end{figure}
\vskip 0.6cm

\subsection{Host Dynamical Masses}

VLT/Keck near-infrared stellar absorption spectroscopy has also been
carried out to constrain the host dynamical mass for many of these
ULIRGs and QSOs. The analysis of our VLT data on ULIRGs (Dasyra et al.
2006ab) builds on the analyses of Genzel et al. (2001) and Tacconi et
al. (2002). We find that the majority of ULIRGs are triggered by
almost equal-mass major mergers of 1.5:1 average ratio, in agreement
with VKS02.  We also find (see Fig. 5) that coalesced ULIRGs resemble
intermediate mass ellipticals/lenticulars with moderate rotation, in
their velocity dispersion distribution, their location in the
fundamental plane (FP; e.g., Kormendy \& Djorgovski 1989) and their
distribution of the ratio of rotation/velocity dispersion [v$_{\rm
  rot}$ sin(i)/$\sigma$]. These results therefore suggest that ULIRGs
form moderate mass ($m^\ast \sim 10^{11}$ M$_\odot$), but not giant (5
-- 10 $\times$ 10$^{11}$ M$_\odot$) ellipticals. These results are largely
consistent with those from our imaging studies.  Converting the host
dispersion in fully coalesced ULIRGs into black hole mass with the aid
of the M$_{\rm BH} - \sigma$ relation (e.g., Gebhardt et al. 2000)
yields black hole mass estimates of the order $10^7 - 10^8$ M$_\odot$.
The accretion rate for sources after the nuclear coalescence is high
$0.5 - 0.9$, again similar to those derived by VKS02 and Veilleux et al.
(2006).  Our preliminary analysis of a dozen PG~QSOs also shows
agreement between the host mass (thus black hole mass) of PG~QSOs and
coalesced ULIRGs (Dasyra et al. 2006c, in prep.).

\vskip 0.3cm

\begin{figure}
\hskip +1.2in
\psfig{file=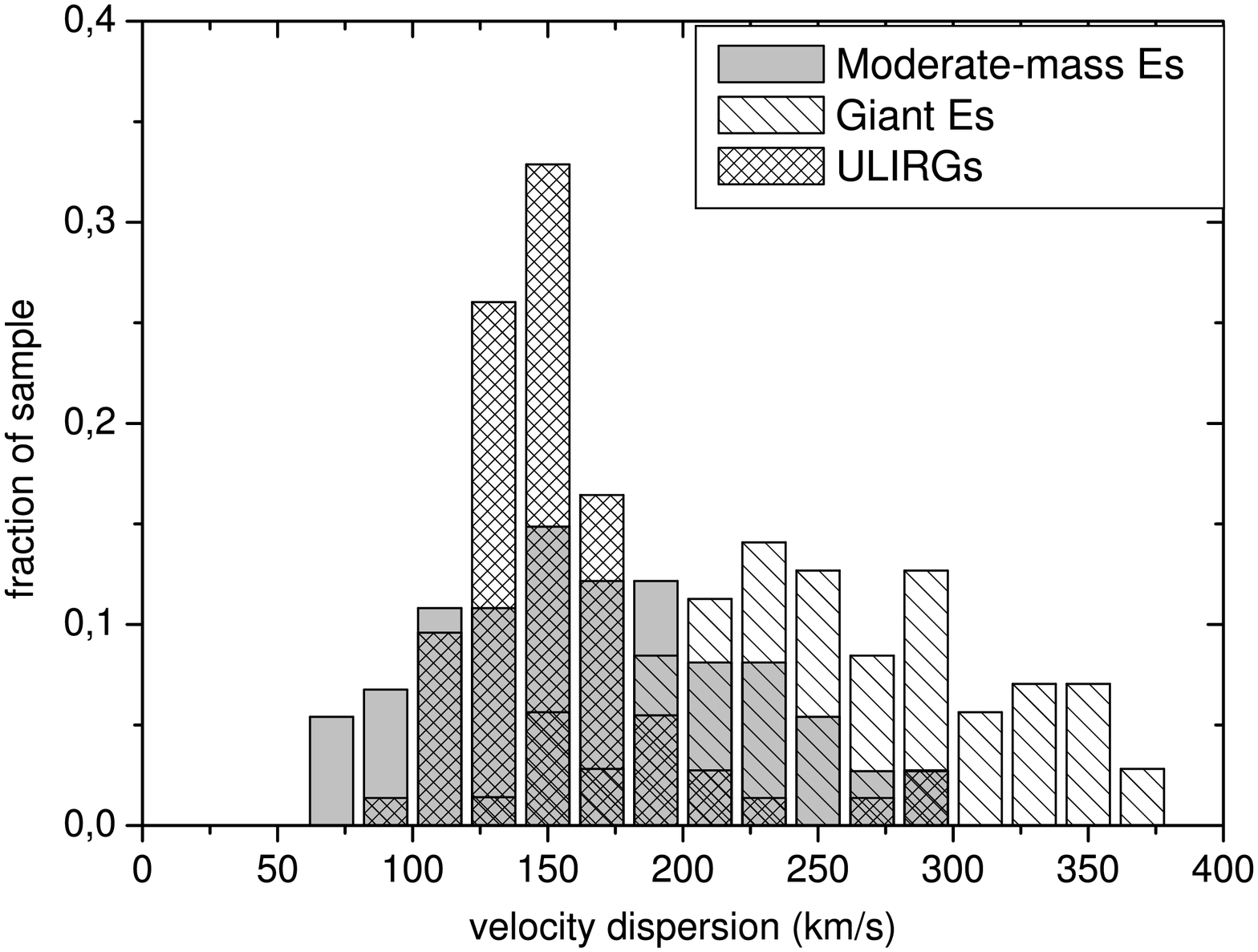,width=3in,angle=-90}
\caption{   Stellar velocity dispersions of ULIRG mergers,
  giant, boxy ellipticals, and intermediate-mass ellipticals and S0s.
  ULIRGs best match intermediate-mass ellipticals. The same is true
  for the ratio of rotation to velocity dispersion (Genzel et al. 2001;
  Tacconi et al. 2002; Dasyra et al. 2006ab).}
\end{figure}
\vskip 0.6cm

\section{Galactic Winds}

As shown in Figure 6, the fraction of objects with detected (neutral
Na~ID-absorbing) winds increases with infrared luminositiy, reaching
$\sim$ 75\% for ULIRGs (e.g., Rupke et al. 2002, 2005; Martin 2005;
see also review by Veilleux, Cecil, \& Bland-Hawthorn 2005).  The
observed detection rate among ULIRGs is consistent with 100\% once
projection effects are taken into account. The projected ``maximum''
velocities in the outflowing components average 300 -- 400 km
s$^{-1}$, and attain $\sim$ 600 km~s$^{-1}$ (although 1100 km~s$^{-1}$
is seen in one object; Fig. 6).  There is some indication of a trend
of increasing outflow velocities with increasing SFR, particularly
when the data from Schwartz \& Martin (2004) on dwarf galaxies are
included.  The winds in ULIRGs entrain considerable neutral material
($\sim$ 10$^8 - 10^{10}$ M$_\odot$ or $\sim$ 10 -- 1000 M$_\odot$
yr$^{-1}$) and are quite powerful ($\sim$ 10$^{56} - 10^{59}$ erg or
10$^{41} - 10^{44}$ erg~s$^{-1}$).  These winds may thus have a
profound effect on the evolution of the ULIRGs. Recent numerical
simulations of mergers with starburst- and AGN-driven feedback provide
a nice theoretical framework for this picture (e.g., Hopkins et al.
2005; Croton et al. 2006).

\vskip 0.3cm

\begin{figure}
\hskip +0.8in
\psfig{file=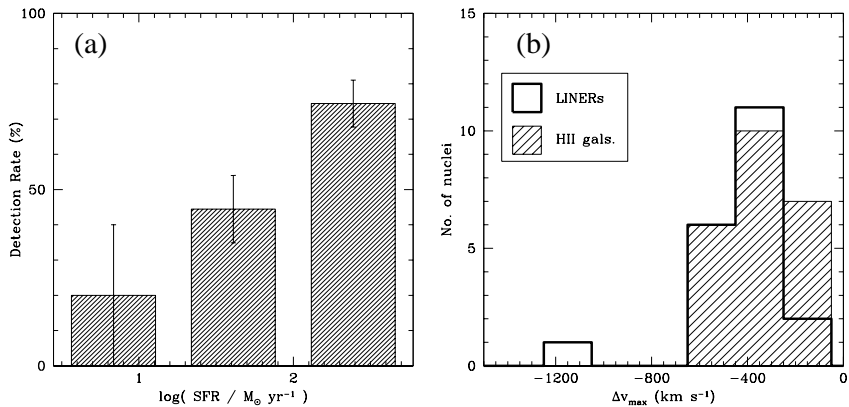,width=4.5in,angle=0}
\caption{ ($a$) Detection rate of winds as a function of star
  formation rate. The last bin on the right corresponds roughly to
  ULIRGs. ($b$) Distribution of maximum outflow velocities among LINER
  ULIRGs (thick line) and H~II region-like ULIRGs (hatched). Adapted
  from Rupke et al. (2005).}
\end{figure}
\vskip 0.6cm

\section{Summary}

The main points of this review can be summarized as follows:

\begin{itemize}
\item[1.] AGN is present and often dominant in $>$ 30\% of all local
  ULIRGs and in $>$ 50\% of local ULIRGs with log~[$L_{\rm
    IR}$/$L_\odot$] $\ga$ 12.3. Optical, near-infrared, and
  mid-infrared diagnostics give consistent answers, but X-ray
  observations at $\la$ 10 keV generally do not.  {\em SST} data under
  analysys will revisit this important issue.  Observations beyond 10
  keV with {\em Suzaku} may also provide interesting constraints.
\item[2.] Warm AGN-like ULIRGs and QSOs have prominent spheroids,
  weaker tidal features, and stronger nuclei than cool LINER or H~II
  ULIRGs. These warm systems are in the final stage of a merger. The
  luminosities, size, and stellar masses of PG~QSOs coincide with
  those of elliptical-like 1-Jy ULIRGs.
\item[3.] Galactic winds are present in virtually all ULIRGs. The
  outflow velocities and the mass and energy injection rates are very
  significant, enough to affect the evolution of ULIRGs. 
\end{itemize}

These results suggests that most ULIRGs are indeed intermediate-mass
ellipticals in formation. Many of these objects may also be in the
process of becoming optical quasars. However, this process is probably
not 100\% efficient: several late-merger ULIRGs do not show obvious
signs of QSO activity.  Moreover, low-luminosity QSOs often harbor
exponential disks which preclude major (1:1 -- 2:1) mergers; this
suggests the existence of a luminosity threshold below which nuclear
activity is not necessarily triggered through major mergers.

\begin{acknowledgements}
  This work was partially funded by NASA through grant GO-0987501A
  from the Space Telescope Science Institute (operated by AURA, Inc.,
  under NASA contract NAS5-26555).  I also thank the organizers, 
  particularly Dr. Yu Gao, for partial financial support, and for
  organizing the meeting in such idyllic surroundings.
\end{acknowledgements}

\label{lastpage}

\end{document}